\documentclass[intlimits,twoside,a4paper]{article}

\usepackage[cp1251]{inputenc}

\usepackage[eqsecnum]{cmpj3}

\usepackage{bm}
\DeclareMathOperator{\sgn}{sgn}

\issue{2020}{23}{1}{13702}
\doinumber{10.5488/CMP.23.13702}

\title[Crossover of spin Hall to quantum anomalous Hall effect in PbC/MnSe heterostructures]{Crossover of spin Hall to quantum anomalous Hall effect in PbC/MnSe heterostructures%
\thanks{Project supported by the National Natural Science Foundation of China (Grant No. 61604106), and Shandong Provincial Natural Science Foundation (Grant No. ZR2014FL025).}}
\author{L. Chen }
\address{School of Physics and Electronic Engineeing, Taishan University, Taian, 271000 Shandong, China
}

\date{Received July 24, 2019, in final form September 26, 2019}

\begin{document}

\maketitle

\begin{abstract}
In this paper, we study topological properties and Hall conductivities  in PbC/MnSe heterostructure under the illumination of a circularly polarized light.  At high frequency regime, energy gap, Chern numbers, and Hall conductivities are studied based on the Floquet theory and Green's function formalism, respectively. The interplay between spin orbit coupling and light leads to  topological phase transition between anomalous Hall states and spin Hall states, which is related to the emission and absorption   of   two virtual photons. The anomalous Hall conductivities are dependent on polarization of light, while the spin Hall conductivity is independent.
\end{abstract}

\section{Introduction}
In the recent years, topological phases and topological phase transitions have been one of the most important topics in modern condensed matter physics~\cite{YRen,HZhaiM}.
The intrinsic anomalous Hall and spin Hall effect have attracted enormous attention in the past decade~\cite{NNagaosa,JSinova}.
The quantum anomalous Hall effect has been theoretically proposed~\cite{LiuCX,YuR}  and
experimentally realized \cite{ChangCZ,KouXF}  in magnetically doped topological insulator  films.
The quantum spin Hall  effect was  theoretically predicted \cite{BABernevig} and experimentally observed~\cite{MKonig} in HgTe quantum wells.
Li {\it et al.} \cite{JianLi} investigated the role of disorder in the topological insulator, which possesses a pair of helical edge states with opposing spins moving in opposite directions and exhibits the phenomenon of quantum spin Hall effect.  The Kane and Mele model defined on a hexagonal lattice can exhibit transition between quantum spin Hall effect phase and a simple insulator due to Rashba spin orbit coupling \cite{CLKane1,CLKane2}.  The  quantum
spin Hall insulator phase has been investigated in the graphene system with a coexistence of Coulomb interaction,
staggered potential, and intrinsic spin-orbit coupling~\cite{JieCao}. In two-dimensional photonic crystal,  a conventional insulator phase, a quantum spin Hall phase, or a quantum anomalous Hall phase can be realized by simply adjusting the geometric parameters and magnetic field \cite{ZGChen,FLiu}. Kort-Kamp \cite{WKortKamp} unveiled topological phase
transitions in the photonic spin Hall effect in the graphene family materials.
The spin Hall effect  can be realized  in a Bose-Einstein condensate of neutral atoms interacting via the magnetic dipole-dipole interactions \cite{TOshima}. A $1$D dynamical version of the quantum spin Hall effect was implemented  in an optical
superlattice with ultracold bosonic atoms \cite{CSchweizer}. Omnidirectional spin Hall effect was studied in a Weyl spin-orbit-coupled atomic gas  \cite{JArmaitis}. The edge state of quantum spin Hall insulators in topological Dirac and Weyl
semimetals has been studied~\cite{XQSun}.
Recently, interaction-driven anomalous quantum Hall state was theoretical predicted~\cite{EVCastro,KSun}.

Another interesting direction of topological states and and topological phase transitions studied in the recent years arises from the nonequilibrium engineering Hall effect under the influence of a periodic drive.
An off resonant circularly polarized light has been found to  induce  a topological term in periodically driven quantum systems to  break the time-reversal symmetry. Based on Floquet theory,
Chen {\it et al.} proposed  a topological quantum phase transition to a quantum anomalous Hall  phase
induced by off-resonant circularly polarized light in a two-dimensional system that is initially in a quantum spin
Hall phase or in a trivial insulator phase \cite{MNChen}.
It has been demonstrated that in monolayer graphene \cite{ZhaiX} and  silicene \cite{EzawaM}, respectively, irradiated by an off-resonance circularly polarized light a topological phase transition between different Hall states takes place.
Laser-induced  quantum anomalous Hall states in honeycomb lattices have also been investigated \cite{Alvaro}.
Weyl semimetals showed a large anomalous Hall effect controllable by illuminating an off-resonant circularly polarized light  \cite{YanZ}.
Previous studies were mostly performed on steady states in
systems with isotropic low-energy dispersions while Hall effects in periodically driven heterostructures  are still lacking.

In this paper,  we use the Floquet theory to study the Hall effects in PbC/MnSe heterostructure~\cite{YLi} with spin orbit coupling under the illumination of an off-resonant circularly polarized light. PbC monolayer grown on an MnSe (111) surface has  a very large energy gap (about 244 meV), and the structure of the system is
mechanically stable. The half-metallic property owned by the heterostructure is protected by the topology of the system and
is robust against the variation of the chemical potential and strain.  Due to these prominent merits,
PbC/MnSe heterostructure provides a novel platform to study the Hall effects.

The structure of this paper is organized as follows. In section~\ref{sec2} we present a brief account of the model and a research method of PbC/MnSe heterostructures with a circularly polarized light, and discuss the evolution of the energy gap. Based on the Floquet theory, we study the Chern numbers and Hall phases in our model at high frequency in section~\ref{sec3}. In section~\ref{sec4}, we study the anomalous Hall and spin Hall conductivities. In section~\ref{sec5}, we summarize our results.

\section{Model and method}\label{sec2}
We begin with the tight-binding Hamiltonian of  PbC/MnSe heterostructure in momentum space on the basis of ${\mid p_{+,\uparrow}\rangle, \mid p_{-,\uparrow}\rangle, \mid p_{+,\downarrow}\rangle,\mid p_{-,\downarrow}\rangle}$ near the Fermi energy around the $\Gamma$ point, which is given by~\cite{YLi}
\begin{eqnarray}
H_0&=&\left({\begin{array}{cccc}
\varepsilon_1(k) &f(k)&0&0\\
f^*(k) & \varepsilon_1(k)&0&0\\
0&0&\varepsilon_1(k)&f(k)\\
0&0&f^*(k)&\varepsilon_2(k)
\end{array}}
\right),
\end{eqnarray}
where $\mid p_{\pm}\rangle=(\mid p_{x}\rangle\pm \ri \mid p_{y}\rangle)/\sqrt{2}$. $\mid p_{x,y}\rangle$ denote the orbital states in Pb, and $\uparrow\downarrow$ represent the
spin-up (-down) state. Taking  time-reversal symmetry and the $C_{3v}$ symmetry into consideration, one can obtain
$f(k_\pm)=\re^{\ri \frac{4\piup}{3}}f(\pm \ri\frac{2\piup}{3}k_\pm)$ with $k_\pm=k_x+\ri k_y$. Therefore, $f(k)$ will take the form $f(k)=\beta k^2_-$, and $\varepsilon_1(k)=\varepsilon_2(k)$ is required. Finally, considering the atomic spin orbit coupling (SOC) interaction,
the $k\cdot p$  Hamiltonian for low-energy physics of the PbC/MnSe heterostructure is
\begin{eqnarray}
\label{TBH}
\hat{H}&=&\alpha k^2+\beta\left[(k_x^2- k_y^2)\hat{\sigma}_x+2k_x k_y\hat{\sigma}_y\right]+s\lambda\hat{\sigma}_z\,,
\end{eqnarray}
where $\alpha$ and $\beta$ are constant taking the  symmetry into consideration. $\hat{\sigma}_{x,y,z}$  denotes the Pauli matrices for the  basis $\mid
p_{\pm}\rangle=(\mid p_{x}\rangle\pm \ri \mid p_{y}\rangle)/\sqrt{2}$.
$\lambda$ denotes SOC interaction  for the spin up ($s=1$) and spin down ($s=-1$) subspace, respectively~\cite{QFLiang}. For simplicity, we take $\beta$ as energy unit and set $\beta=1$ throughout the paper.

\begin{figure} [!t]
\begin{center}
 \includegraphics[width=8.0cm]{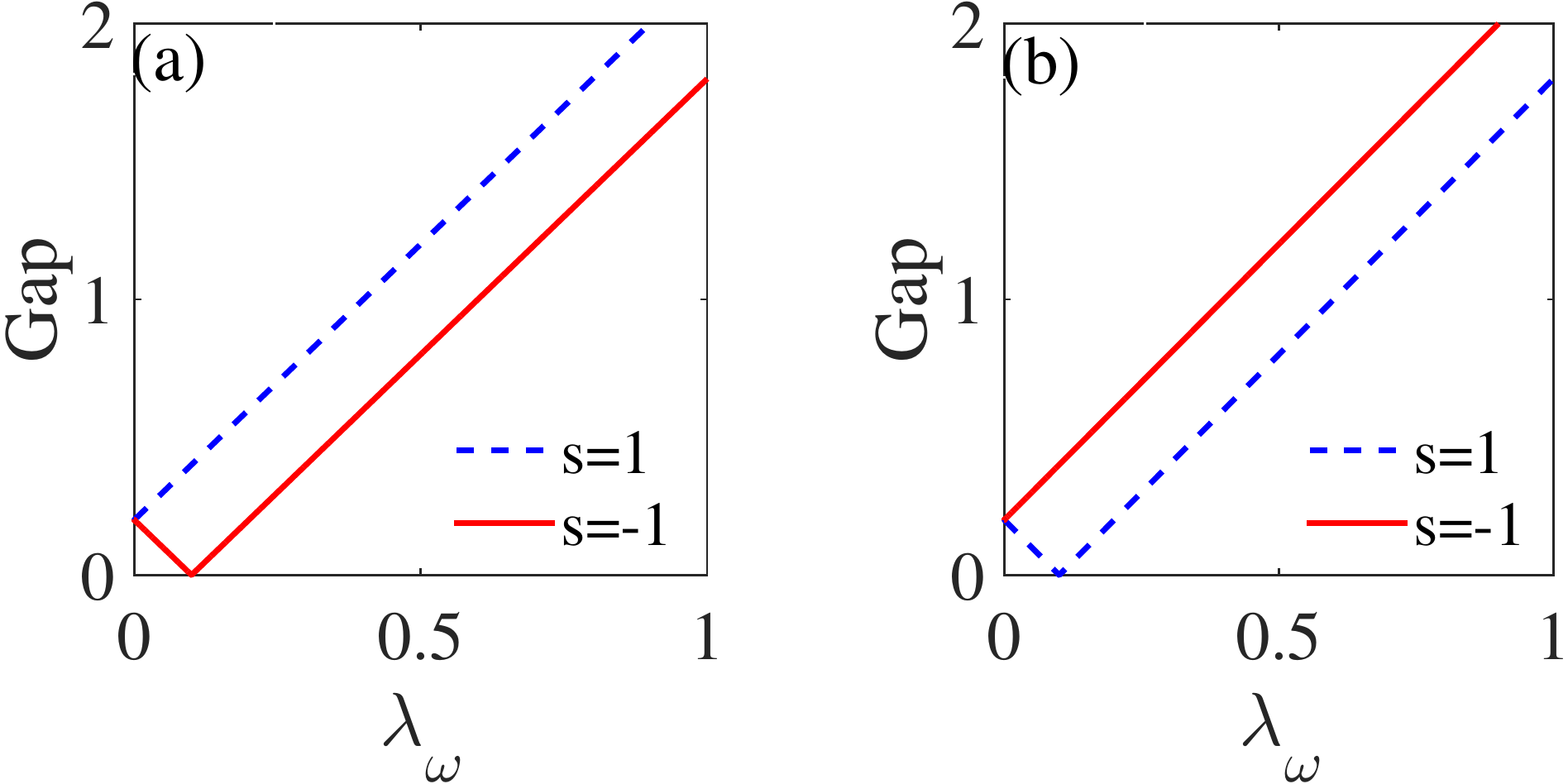}
 \end{center}
\caption{ (Colour online) Bulk band gap as a function of the light induced term
$\lambda_{\omega}=\Delta_{\omega}|_{k=0}$ when (a) $\kappa=1$ and (b) $\kappa=-1$. The blue dashed and red solid lines denote spin-up and spin-down bands, respectively.
$\lambda$ is set to be $0.1$.}\label{fig1}
 \end{figure}
Considering the  circularly polarized light irradiated onto the PbC/MnSe heterostructure, the  vector  potential   introduced  into  the  Hamiltonian  by
substitution $\vec{k}\rightarrow\vec{k}+e\vec{A}(t)$ is as follows:
\begin{eqnarray}
\vec{A}(t)=A(\kappa\sin{\omega t},\cos{\omega t}),
\end{eqnarray}
where $\omega$ is the frequency of light, $\kappa=\pm1$ denote right-hand and left-hand polarization of light, respectively. $A$ is the amplitude of the
vector potential. Based on Floquet theory~\cite{TOka,TKitagawa,NHLindner}, the photon dressed effective Hamiltonian with weak  light  intensity can be
written as follows:
\begin{eqnarray}
\hat{H}^\text{F}=\hat{H}+\frac{[\hat{H}_{-1},\hat{H}_{1}]}{\hbar\omega}+\frac{[\hat{H}_{-2},\hat{H}_{2}]}{2\hbar\omega}\,,
\end{eqnarray}
where
\begin{eqnarray}
\hat{H}_{\pm 1}&=&\frac{\omega}{2\piup}\int_{0}^{\frac{2\piup}{\omega}}\hat{H}(t)\re^{\pm \ri\omega t},\\
\hat{H}_{\pm 2}&=&\frac{\omega}{2\piup}\int_{0}^{\frac{2\piup}{\omega}}\hat{H}(t)\re^{\pm 2\ri\omega t}.
\end{eqnarray}
$\hat{H}(t)$ is the time-dependent  Hamiltonian using the substitution $\vec{k}\rightarrow\vec{k}+e\vec{A}(t)$ in equation~(\ref{TBH}). Straightforward calculations show
\begin{eqnarray}\label{LEH}
\hat{H}^\text{F}&=&\alpha k^2+\big(k_x^2- k_y^2\big)\hat{\sigma}_x+2k_x k_y\hat{\sigma}_y
+[s\lambda+\kappa\Delta_{\omega}]\hat{\sigma}_z\,,
\end{eqnarray}
where
\begin{eqnarray}\label{Egap}
\Delta_{\omega}=\frac{1}{\hbar\omega}\left(\frac{eA}{\hbar}\right)^2\left[4k^2+\frac{1}{2}\left(\frac{eA}{\hbar}\right)^2\right].
\end{eqnarray}
We note that an extra light induced  term ($\Delta_{\omega}$) is introduced, and its sign depends on $\kappa$ rather than on spin.
This new term is clearly different from the intrinsic SOC term ($s\lambda$) and also different from the conventional Zeeman term.
For $\lambda>0$ and $\Delta_{\omega}<0$, the overall value of the effective SOC ($s\lambda+\kappa\Delta_{\omega}$) will
decrease (increase) if $s$ and $\eta$ have the same (different) sign. The critical point $\lambda_{\omega}$ is defined by $s\lambda+\kappa\Delta_{\omega}=0$ at $k=0$,
where the  $\lambda_{\omega}$ vanishes and band gap closes. Consequently, the spin-up and -down bands will
have an opposite response to the circularly polarized light of the opposite $\kappa$. Figure~\ref{fig1}~(a) shows the gap  as a function of the light induced
term $\lambda_{\omega}=\Delta_{\omega}|_{k=0}$ for $\kappa=1$ and $\lambda=0.1$. When  increasing $\lambda_{\omega}$, the  gap of spin-down component  decreases while that of the spin up component increases. Then, the spin-down gap closes and reopens. Figure~\ref{fig1}~(b) shows that similar  changes are found for $\kappa=-1$.
It indicates that a topological phase transition takes place. The topologically different responses  arise from the interplay between the SOC and light.

\section{Chern numbers and Hall phase}\label{sec3}
 For our system, the Hamiltonian can always be expressed  in terms of the Pauli matrices as follows:
\begin{eqnarray}\label{hamdv}
\hat{H}^\text{F}({\bf k})={\bf d}({\bf k})\cdot\hat{\sigma}+\epsilon({\bf k})\mathbf{I}\,,
\end{eqnarray}
where ${\bf d}({\bf k})=(d_{x}({\bf k}),d_{y}({\bf k}),d_{z}({\bf k}))$, $\hat{\sigma}=(\sigma_{x},\sigma_{y},\sigma_{z})$ , and $\mathbf{I}$ is
the rank-2 unit matrix.
From the above effective Hamiltonian, the Chern number is calculated as~\cite{XLQi}
\begin{eqnarray}
C(s)=\frac{1}{2\piup}\int_{BZ}\rd^2\textbf{\emph{k}}\Omega(s)\,,
\label{ChernN}
\end{eqnarray}
where $\Omega(s)$ is the  Berry curvature  in the momentum space over all occupied states
of electrons with spin $s$ component, and it can be written as follows:
\begin{eqnarray}
\Omega(s)=\frac{1}{2}\hat{d}\cdot\left(\frac{\partial\hat{d}}{\partial k_x}\times\frac{\partial\hat{d}}{\partial k_y}\right),
\label{bc}
\end{eqnarray}
where $\hat{d}=\frac{{\bf d}}{|{\bf d}|}$. Straightforward calculation shows
\begin{eqnarray}\label{bcre}
\Omega(s)=\frac{2k^2\left[s\lambda+\kappa\frac{1}{2\hbar\omega}\left(\frac{eA}{\hbar}\right)^4\right]}
{\left[k^4+(s\lambda+\kappa\Delta_{\omega})^2\right]^{3/2}},
\end{eqnarray}
and
\begin{eqnarray}\label{CN}
C(s)=\sgn(s\lambda+\kappa\lambda_\omega)\,,
\end{eqnarray}
where $\sgn$ is the sign function. One can see that the Chern number will change its sign with the varying of $\sgn(s\lambda+\kappa\lambda_\omega)$.

For $\alpha=\lambda=0$, the Hamiltonian [equation~(\ref{LEH})] is similar to the bilayer graphene irradiated with circularly polarized light~\cite{SMorell,LAbergela,IVIorsh,ChunleiQ}. In the bilayer graphene irradiated with circularly polarized light, the light induces a term which is valley and polarization dependent. Therefore, the Berry curvatures are the valley and polarization dependent, and the integration gives an integer nonzero Chern number per valley, hence realizing  a quantum valley-Hall state. In our system,  a quantum spin Hall state probably exists due to SOC.

\begin{figure} [!b]
\begin{center}
\includegraphics[width=5.0cm]{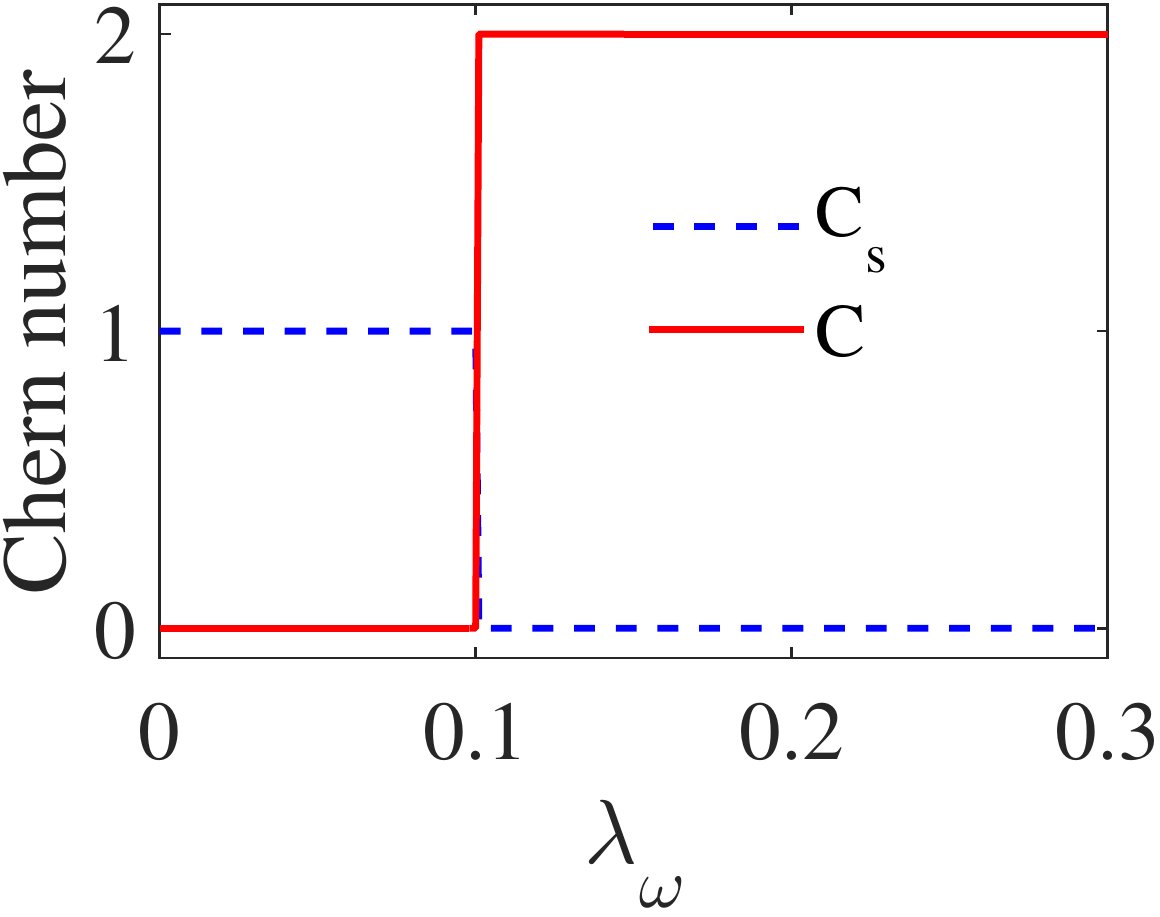}
 \end{center}
\caption{ (Colour online) \small{ Chern numbers as a function of the light induced term
$\lambda_{\omega}=\Delta_{\omega}|_{k=0}$   ${C}_s$  (dashed blue line) and  $C$ (solid red line) for $\lambda=0.1$ and $\kappa=1$.}}\label{chernn}
 \end{figure}

 In figure~\ref{chernn}, we plot Chern numbers as a function of $\lambda_{\omega}=\Delta_{\omega}|_{k=0}$ for $\lambda=0.1$ and $\kappa=1$. When the Fermi level lies inside the bulk energy gap, it can be found that the quantum state is  a quantum spin Hall (QSH) insulator when there is no  circularly polarized light is absent, which is characterized by a nonzero spin Chern number ${C}_s=\frac{1}{2}[C(1)-C(-1)]$. We have considered that the spin $s_z$ is a good quantum number.
With an increasing circularly polarized light intensity, the spin Chern number ${C}_s$ changes from $\sgn(\lambda)$ to $0$, while
the Chern number $C=C(1)+C(-1)$ changes from $0$ to $2\kappa$ with closure  and reopening of the gap. The  nonzero Chern number $C$ means that the insulating state is a quantum anomalous Hall (QAH) insulator. Therefore,
a QSH to QAH phase transition takes place, which is consistent with our numerical results that the gap closes and reopens. It should be noted that  the topological phase transition results from emission and absorption of two virtual photons (see equation~\ref{Egap}) in this paper.
In the next section we will study the  anomalous Hall and spin Hall conductivities to confirm our results. Thus, there is a direct correspondence between the Chern number and the Hall conductance for two-dimensional  insulators  ($\sigma_{xy}=Ce^2/h$) when the Fermi level lies inside the bulk energy gap.

It should be noted that the first term in equation~(\ref{LEH}) in the revised manuscript  does not modify the eigenstates of our system.  Therefore, we will set $\alpha=0$ to explore the Hall effects in the remainder of this paper. Furthermore,
the critical point is ${e}A/{\hbar}=(2s\kappa\lambda\omega)^{\frac{1}{4}}$ when the topological phase transition take place based on the Hamiltonian~(\ref{LEH}).  The atomic spin orbit is small which is responsible for the opening of the gap. At high frequency regime, we have set $\omega=6$. Thus the light intensity ${e}A/{\hbar}$ can be tuned to be small, which can be realized experimentally. On the other hand, the number of the edge state is characterized by Chern number, and topological phase transition is associated with the closure and reopening of the energy of the gap. Thus, we can observe the closure and reopening of the energy, the number of each edge, and propagation direction of the each edge
to confirm the topological phase transition experimentally.

\section{Anomalous Hall and spin Hall conductivities}\label{sec4}
To better understand the topological properties, we studied the anomalous Hall and Spin Hall conductivities  in this section.
Based on Green's function theory \cite{OnodaS,KovalevAA}, anomalous Hall and spin Hall can be expressed as follows:
\begin{eqnarray}
  \sigma_{xy}&=&\sigma_{xy}^{c \text I}+\sigma_{xy}^{c \text {II}},\\
  \sigma_{xy}^{s_z}&=&\sigma_{xy}^{s\text {I}}+\sigma_{xy}^{s \text {II}}.
\end{eqnarray}
The first part is the contribution from the Fermi surface, the other
part is the contribution from the Fermi sea, and
\begin{eqnarray}
  \sigma_{xy}^\text{c(s)\text {I}} &=& -\frac{e\hbar}{2}\int\frac{\rd\varepsilon}{2\piup}\partial_\varepsilon f(\varepsilon)
  \int\frac{\rd^2\textbf{k}}{(2\piup\hbar)^2}
{\rm Tr}
  \left[\hat{j}_\text{c(s)}(\textbf{k})\hat{G}_0^\text R(\varepsilon,\textbf{k})\hat{v}_y(\textbf{k})
  \big(\hat{G}_0^\text A(\varepsilon,\textbf{k})-\hat{G}_0^\text R(\varepsilon,\textbf{k})\big)
   \right.\nonumber\\
    &&\left.-\hat{j}_\text {c(s)}(\textbf{k})\big(\hat{G}_0^\text A(\varepsilon,\textbf{k})-\hat{G}_0^\text R(\varepsilon,\textbf{k})\big)\hat{v}_y(\textbf{k})
  \hat{G}_0^\text R(\varepsilon,\textbf{k})
  \right],
  \label{shesigma^I}\\
  \sigma_{xy}^\text{c(s)\text {II}} &=& -\frac{e\hbar}{2}\int\frac{\rd\varepsilon}{2\piup}  f(\varepsilon)
  \int\frac{\rd^2\textbf{k}}{(2\piup\hbar)^2}
{\rm Tr}
  \left[\hat{j}_\text{c(s)}(\textbf{k})\hat{G}_0^\text A(\varepsilon,\textbf{k})
  \hat{G}_0^\text A(\varepsilon,\textbf{k})\hat{v}_y(\textbf{k})\hat{G}_0^\text A(\varepsilon,\textbf{k})
     \right.\nonumber\\
    &&\left.-\hat{j}_\text{c(s)}(\textbf{k})\hat{G}_0^\text A(\varepsilon,\textbf{k})
\hat{v}_y(\textbf{k})\hat{G}_0^\text A(\varepsilon,\textbf{k})  \hat{G}_0^\text A(\varepsilon,\textbf{k})
-\big(\hat{G}_0^\text A\rightarrow\hat{G}_0^\text R\big)\right],
  \label{shesigma^II}
\end{eqnarray}
where $f(\varepsilon)$ is the Fermi distribution. $\hat{j_\text c}=\hat{v}_x$ and $\hat{j_\text s}=\frac{1}{2}\{\hat{v}_x,\hat{s}_{z}\}$ are the charge and spin current density operator, respectively.  $\hat{v}_{x,y}=(1/\hbar)(\partial H/\partial p_{x,y})$ is the velocity operator.
\begin{equation}
  \hat{G}^\text{R,A}_0(\varepsilon,\textbf{k})=[\varepsilon-\hat{H}^\text F(\textbf{k})
\pm \ri 0^+]^{-1}
  \label{eq:G^R,A:0}
\end{equation}
is the retarded (advanced) Green's function in the clean limit.

\begin{figure} [tbp]
\begin{center}
 \includegraphics[width=8.0cm]{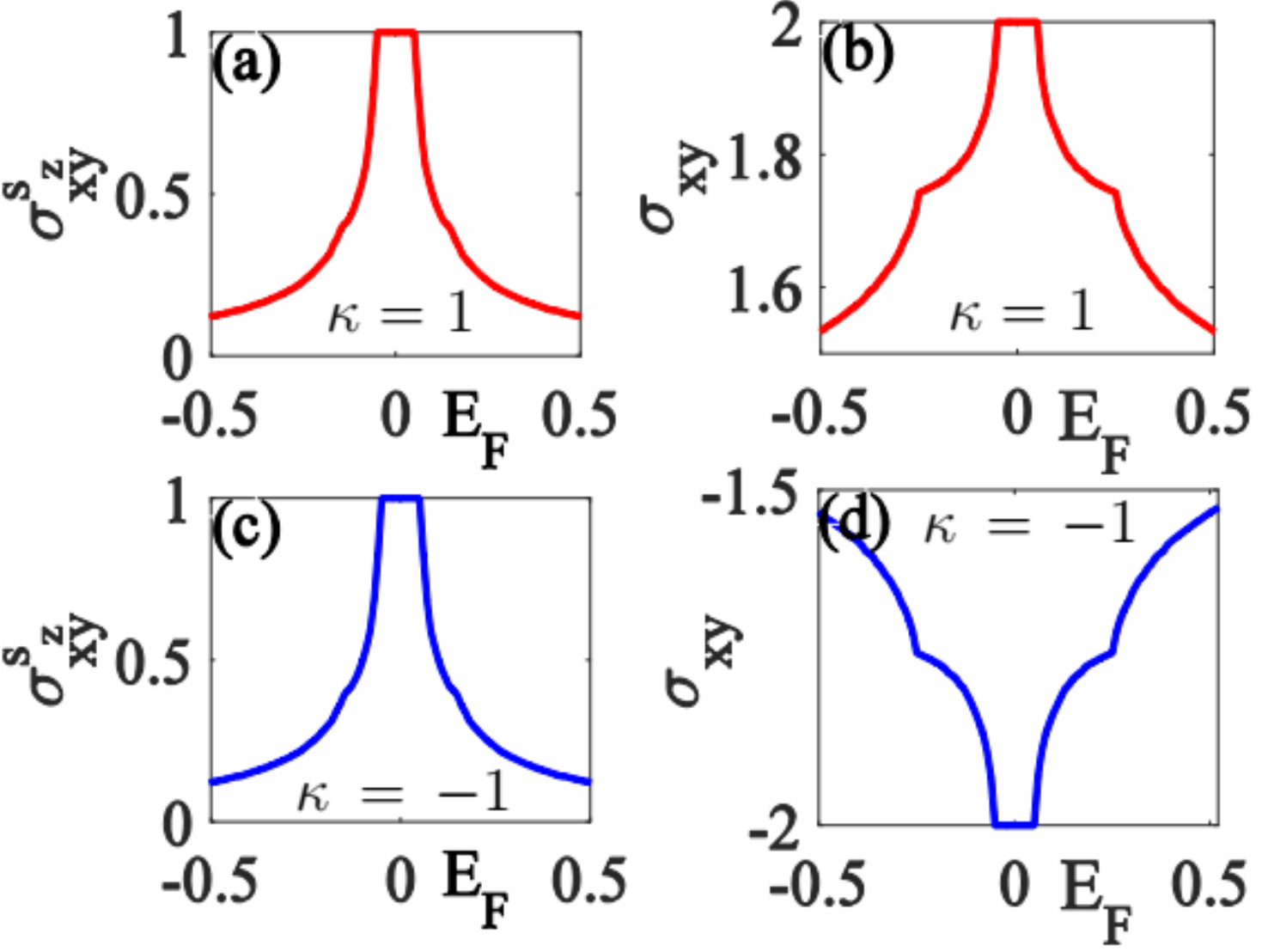}
 \end{center}
\caption{ (Colour online) \small{Spin Hall and Hall conductivity in units of $e/2\piup$ and $e^2/2\piup\hbar$, respectively, under the irradiation of circularly polarized light. The red and blue solid lines denote the right-hand and left-hand circularly polarized light, respectively. The typical parameters $\lambda_{\omega}$ are set to $0.05$ in (a) (c) and $0.15$ in (b) (d).  }}\label{fig2}
 \end{figure}

Figure~\ref{fig2} shows numerical results of the  spin Hall ($\sigma^{s_z}_{xy}$) and anomalous Hall conductivities ($\sigma_{xy}$) for both the left-hand and right-hand circularly polarized lights.
As shown in figures~\ref{fig2} (a) and (c), the system is QSH insulator with quantized spin Hall conductivity of $e/2\piup$ when the Fermi level lies inside the energy gap for both circularly polarized lights before the closure of the energy  gap. However,  after reopening of the  energy gap, the spin Hall conductivity is quenched to zero. This is because the Chern number for both two-spin components is determined by the sign of  $s\lambda+\kappa\lambda_{\omega}$ (see equation~\ref{CN}). We can also find that the Chern number $C=2$ and $C=-2$ for the QAH state under the left-hand and right-hand circularly polarized lights, respectively, after reopening the energy gap. This is consistent with the quantized anomalous Hall conductivity of $e^2/2h$ and $-e^2/2h$  for the two circularly polarized lights, respectively, as shown in figures~\ref{fig2} (b) and (d). Consequently, a QSH to QAH topological phase transition is realized, and it depends on the polarization of light.

On the other hand, $\sigma^{s_z}_{xy}$ [figures~\ref{fig2} (a) and (c)] drops  promptly from the quantized plateau. This behavior is independent of polarization of light. When the Fermi energy ($|E_\text F|$) is outside the energy gap, the $\sigma^{s_z}_{xy}$ decreases with increasing $|E_\text F|$. The $\sigma_{xy}$ [figures~\ref{fig2} (b) and (d)]  is drops/raises promptly from the quantized plateau for the right-hand/left-hand circularly polarized light. When $|E_\text F|$ is outside the energy gap, the $|\sigma_{xy}|$ decreases with increasing $|E_\text F|$. There is a link in $\sigma_{xy}$.
The nonzero Berry curvature of the band mainly distributes near its band edge, leading to this behavior of $\sigma^{s_z}_{xy}$ and $\sigma_{xy}$.
Based on equation~(\ref{bcre}), $\sigma^{s_z}_{xy}$ and $\sigma_{xy}$ will always vanish at a higher energy.
\section{Conclusion and summary}\label{sec5}
To conclude, we have analyzed a topological phase transition and Hall conductivities  in PbC/MnSe heterostructure under the illumination of a circularly polarized light. We have employed Floquet theory and Green's function formalism to compute Chern numbers and Hall conductivities at high-frequency regime, respectively. The Chern numbers are determined by the interplay between SOC and light induced SOC-like term. A photo-induced transition between quantum spin Hall state and anomalous Hall state was observed. When the Fermi energy lies in the energy gap, the anomalous Hall conductivities are dependent on polarization of light, while spin Hall conductivity is independent. Our findings provide a way to control the topological phase transition and transport properties of the heterostructure. This is also important for studying the effect of spin-orbit coupling in a heterostructure.



%
%

\newpage
\ukrainianpart

\title{Кросовер спін-холлівського ефекту до квантового аномального ефекту Холла в  PbC/MnSe гетероструктурах}
\author{Л. Чен}
\address{Вища школа фізики та електроніки, університет Тайшань, 271000 Шаньдун, Китай
}

\makeukrtitle

\begin{abstract}
	У статті  досліджуються топологічні властивості та холлівські провідності  у гетероструктурі PbC/MnSe при освітленні циркулярно поляризованим світлом. У високочастотному режимі досліджуються заборонена зона, числа Черна та холлівські провідності на основі теорії Флоке і формалізму функцій Гріна, відповідно. 
	Взаємозв'язок між спін-орбітальною взаємодією і світлом приводить до топологічного фазового переходу між аномальними станами Холла та спін-холлівськими станами, що, в свою чергу, пов'язано з випромінюванням і поглинанням двох віртуальних фотонів.
	Аномальні провідності Холла залежать від поляризації світла, в той час як спін-холлівська провідність є незалежною.
	\keywords Холл, світло, теорія Флоке, топологія
\end{abstract}

\end{document}